\documentclass[twocolumn,showpacs,preprintnumbers,amsmath,amssymb,eqsecnum]{revtex4}

\usepackage{graphicx}
\usepackage{amssymb}
\usepackage{amsthm}
\usepackage{dcolumn}
\usepackage{bm}

\def\pp{\partial}

\begin{document}

\bibliographystyle{unsrt}

\title{Stabilization of ultra-short pulses in cubic nonlinear media}

\author{Y. Chung$^a$ and T. Sch{\"a}fer$^b$}

\affiliation{$^a$ Department of Mathematics,
                Southern Methodist University,
                Dallas, TX 75275, USA  \\
$^b$ Department of Mathematics,
          College of Staten Island,
	  City University of New York,
          Staten Island, NY 10314, USA}
\date{\today}


\begin{abstract}
We study the propagation of ultra-short pulses in a cubic nonlinear
medium. Using multiple-scale technique, we derive a new wave equation
that preserves the nonlocal dispersion terms present in Maxwell's
equations. As a result, we are able to understand how ultra-short
nonlinear shocks are stabilized by dispersive terms. A delicate
balance between dispersion and nonlinearity leads to a new type of
solitary waves. Their stability is confirmed by numerical simulations
of full Maxwell's equations.
\end{abstract}

\maketitle

\section{Introduction}
Recent progress in nonlinear optical systems has led to increasing
development and use of ultra-short technologies. Driven by the advent
of novel detection techniques such as frequency resolved optical
gating (FROG), the precise measurement of ultra-short pulses is now
possible. Understanding the dynamics of ultra-short pulses has been
the subject of intensive research due to their potential uses for
various fields, e.g., medical applications and imaging \cite{bwzw99},
continuum recompression and control from highly nonlinear photonic
crystal fibers \cite{ktrs00, k98}, and the next generation of
telecommunication systems.

The cubic nonlinear Schr{\"o}dinger equation (NLSE) has been the
governing model for pulse propagation in nonlinear medium. The NLSE,
however, was shown to be inadequate for describing ultra-short
pulses. As the pulse power increases, i.e., the pulse becomes shorter,
the delay in the response of the media to the excitation of the
electric field starts to play a dominant role. The characteristics of
the delay are not instantaneous, and hence they need to be described
by convolution terms in governing equations. In contrast to
differential operators, which are local operators, convolution terms
involve nonlocal operators and hence the resulting system is an
integro-differential equation. The NLSE is a local equation, namely,
it only involves local terms approximating the delayed response. Thus,
the NLSE remains valid only while the basic assumption, the existence
of fairly broad pulses, is guaranteed \cite{rothenberg:1992}.

Current optical technology allows to design state-of-the-art optical
devices such as photonic crystals or microstructured fibers whose
structures are more complex than standard optical fibers. Experimental
observations have shown that these new devices provide remarkable
phenomena never seen in standard optical fibers. Some of the most
prominent phenomena include creation of photonic band gaps,
supercontinuum generation with relatively low pulse intensities,
ultrashort pulse generation, pulse fission reminiscent of multisoliton
breakup \cite{husakou, hgz02, dudley}, and the simultaneous third
harmonic generation \cite{fio}. While photonic crystal technology can
successfully demonstrate the remarkable performance of the devices,
very little is known in theoretical study. It is evident that one must
develop a reliable mathematical model that can explain these new
optical phenomena. In fact, any envelope approximation that eventually
leads to the cubic NLSE fails to exhibit the optical mechanism which
evidently goes beyond the local soliton dynamics.

A considerable amount of research efforts have been made to capture
the complex optical phenomena that cannot be explained by the cubic
NLSE. For short pulse dynamics, for example, one needs to introduce
higher order dispersive terms, which lead to modified NLSE whose
structure, although much more complex, still remains local. For
certain materials, e.g. bulk silica, one can derive a local equation
which describes the evolution of ultra-short pulses by making use of
the specific form of the material susceptibility, $\chi^{(1)}$
\cite{schaefer-wayne:2004}. Recently, the short pulse equation (SPE)
was independently derived, which was proven to be a better
approximation as the NLSE loses accuracy
\cite{chung-jones-etal:2005}. The SPE, however, was not shown to
possess soliton solutions \cite{schaefer-wayne:2004}. Moreover, the
SPE is not valid for all materials since the equation is derived
taking into account a particular type of the susceptibility of bulk
silica. In this paper, considering general forms of susceptibility, we
derive a new model without imposing any locality assumption and
therefore, avoiding potential loss of smoothness that could have
helped to obtain stable solutions in SPE. While soliton theory has
been developed in the context of locality, it is now our main task to
investigate if nonlocal equations can provide similar phenomena.
Indeed, we show that our new nonlocal equation also possesses solitary
waves. Furthermore, we perform numerical simulations to confirm the
stability of the solutions, which suggests that the soliton theory can
be extended to systems with convolution terms.

The paper is organized as follows. In section \ref{derivation}, we
present the derivation of the nonlinear nonlocal wave equation from
Maxwell's equations. In section \ref{multiple}, using multiple-scale
technique, we derive a weakly nonlinear approximation to the wave
equation introduced in section \ref{derivation}. In section
\ref{dispersion-free}, we show that the absence of dispersion leads to
the shock formation. In section \ref{dispersion}, we further stabilize
the optical shocks by introducing a dispersive convolution term, and
show our model possesses solitary waves. In section \ref{numerics}, we
confirm the stability of the solitary wave solutions by numerical
simulations. Finally, in section \ref{summary} we summarize the main
results of our analysis.
 
\section{Derivation of nonlinear nonlocal wave equation}\label{derivation}

Prominent examples of effects that lead to convolution terms are
optical phenomena due to the retardation in the response of the media
to excitation by an oscillating electric field
\cite{boyd:1992}. Recently, the relation between nonlocal terms and
quadratic solitons in quadratic nonlinear $\chi^{(2)}$-materials has
been studied \cite{nikolov-neshev-etal:2003}. Here, we consider cubic
nonlinear $\chi^{(3)}$ materials and investigate their impacts on wave
propagation. We start with Maxwell's equations:
\begin{eqnarray}\label{maxwell}
&&\nabla\times {\mathbf{E}} = -\frac{\partial {\mathbf{B}}}{\partial t}, \qquad \nabla\cdot {\mathbf{D}} = \rho, \nonumber\\
&&\nabla\times {\mathbf{H}} = {\mathbf{j}} + \frac{\partial {\mathbf{D}}}{\partial t}, \qquad \nabla \cdot {\mathbf{B}} = 0.
\end{eqnarray}
The fields ${\mathbf{E}}$, ${\mathbf{B}}$, ${\mathbf{D}}$,
${\mathbf{H}}$ are real-valued vectors depending on the three space
variables $x$, $y$, $z$ and the time $t$; $\rho$ is the charge density
and ${\mathbf{j}}$ is the current density vector. We write $\Phi_1$,
$\Phi_2$, $\Phi_3$ to denote the the first, second, and the third
component of the vector ${\mathbf{\Phi}}$.
The main results of this paper do not depend on the dimension of the
considered system. Thus, for simplicity, the analysis is carried out in an one-dimensional context. This allows us to reduce the Maxwell's equations to {\em one} dimensional system of equations by setting
\begin{equation}
{\mathbf{E}}=E_3(x,t)\vec{\mathbf{e}}_3, \qquad {\mathbf{H}} = H_2(x,t)\vec{\mathbf{e}}_2.
\end{equation}
In dielectric media, there exist neither free charges nor field sources, and hence we can set ${\mathbf{j}}=0$ and $\rho=0$. We also assume ${\mathbf{B}}=\mu_0 {\mathbf{H}}$ where
$\mu_0$ is a constant, which is appropriate for bulk silica. The relation
between ${\mathbf{D}}$ and ${\mathbf{E}}$ is given by
\begin{equation}
D_3 = \epsilon_0 E_3 + P_3
\end{equation}
where $\epsilon_0$ is the permeability and $P_3$ will be defined
later. These assumptions reduce the system of equations
(\ref{maxwell}) to two one-dimensional equations:
\begin{equation}
\frac{\partial H_2}{\partial x}=\frac{\partial D_3}{\partial t}, \qquad \frac{\partial E_3}{\partial x}=\mu_0\frac{\partial H_2}{\partial t}.
\end{equation}
Using the above relations, we find the equation for $E_3$, 
\begin{equation}
\frac{\partial^2E_3}{\partial x^2}-\epsilon_0\mu_0\frac{\partial^2E_3}{\partial t^2}=\mu_0\frac{\partial^2P_3}{\partial t^2}.
\end{equation}
Now we set $c^2=(\mu_0\epsilon_0)^{-1}$ and obtain the
standard form of the wave equation:
\begin{equation}
\frac{\partial^2E_3}{\partial x^2}-\frac{1}{c^2}\frac{\partial^2E_3}{\partial t^2}=\frac{1}{c^2}\frac{\partial^2}{\partial t^2}\left(\frac{P_3}{\epsilon_0}\right).
\end{equation}
Since the response of the medium is not instantaneous, we account for this
{\em retarded response} in terms of convolution integrals and obtain 
\begin{widetext}
\begin{eqnarray}
P_3 = \epsilon_0 \int \chi^{(1)}(t-\tau)E_3(x,\tau)d\tau + \epsilon_0\int \chi^{(3)}(t-\tau_1,t-\tau_2,t-\tau_3)E_3(x,\tau_1)E_3(x,\tau_2)E_3(x,\tau_3)d\tau_1d\tau_2d\tau_3.
\end{eqnarray}
\end{widetext}
Note that this is where convolution terms enter naturally. We finally
find one dimensional nonlinear wave equation (replacing $E_3$ by $E$
and setting $c=1$)
\begin{widetext}
\begin{eqnarray} 
\frac{\partial^2E}{\partial x^2}=\frac{\partial^2E}{\partial t^2}+\frac{\partial^2}{\partial t^2}\left(\int \chi^{(1)}(t-\tau)E(x,\tau)d\tau 
 + \int \chi^{(3)}(t-\tau_1,t-\tau_2,t-\tau_3)E(x,\tau_1)E(x,\tau_2)E(x,\tau_3)d\tau_1 d\tau_2 d\tau_3\right)\label{nonlinwave}
\end{eqnarray}
\end{widetext}
Here, convolution terms are present in both the dispersive and the
nonlinear terms. Generally, the convolution integral in the dispersion
has more significant impact on pulse propagation. The Fourier
transform helps to convert the convolution present in the dispersive
term to a product, however, even if the nonlinearity was local, it
would lead to a convolution of the nonlinear term in Fourier domain.

\section{Nonlocal terms in multi-scale expansions}\label{multiple}

Multiple-scale technique allows to derive asymptotic
expansions for ordinary and partial differential equations
\cite{holmes:1995}. The basic idea is to use the slow scales to remove 
resonances which are responsible for short validity range of the chosen expansion. In this paper, we focus on the case where dispersion and
nonlinearity effects are weak and of the same order $\epsilon$. The method presented here is very general and hence, can be applied to other cases. In the presence of weak dispersion and nonlinearity, the
calculations are more straightforward to perform since the zero-order operator is the unperturbed linear wave equation whose kernel can be constructed by
simple exponential functions. This makes it easy to apply the Fredholm
alternative leading to the solvability conditions which result in the
equations for the slow evolution of the system.

Let us write the basic wave equation again, this time with a small
dispersive and small nonlinear term on the right hand side:
\begin{widetext}
\begin{eqnarray} 
\frac{\partial^2E}{\partial x^2}= \frac{\partial^2E}{\partial t^2}+
\epsilon\frac{\partial^2}{\partial t^2}\left(\int \chi^{(1)}(t-\tau)E(x,\tau)d\tau
 + \int \chi^{(3)}(t-\tau_1,t-\tau_2,t-\tau_3)E(x,\tau_1)E(x,\tau_2)E(x,\tau_3)d\tau_1 d\tau_2 d\tau_3\right) \label{maxwell_disp_weak}
\end{eqnarray}
\end{widetext}
We introduce multiple scales {\em only} in the
evolution variable $z$, and thus no additional scales in $t$. Then, in contrast to the derivation of the NLSE, our
analysis is valid even in the case where the Fourier transform is not
concentrated around one special frequency. This is of particular
importance for the study of very short pulses or phenomena involving
frequency-mixing.
The first step is to rewrite (\ref{maxwell_disp_weak}) in Fourier
domain
\begin{equation}\label{maxwell_disp_weak_fourier}
\left(\frac{\partial^2}{\partial x^2}+\omega^2\right)\hat{E} = -\epsilon\omega^2\hat{\chi}^{(1)}(\omega)\hat{E}(x,\omega) 
-\epsilon\frac{\omega^2}{(2\pi)^2} {\mathcal{N}}(E), 
\end{equation}
where we have introduced the nonlinear (and nonlocal) operator
\begin{eqnarray}
{\mathcal{N}}(E)&=&\int \hat\chi^{(3)}(\omega_1,\omega_2,\omega_3)\hat{E}(x,\omega_1)\hat{E}(x,\omega_2)\hat{E}(x,\omega_3)\nonumber\\
&\times&\delta(\omega-\omega_1-\omega_2-\omega_3)\;d\omega_1\;d\omega_2\;d\omega_3. 
\end{eqnarray}
Now we perform a multiple scale expansion
\begin{eqnarray}
&&\hat{E}(x,\omega)=\hat{E}_0(x_0,x_1,\omega)+\epsilon \hat{E}_1(x_0,x_1,\omega) +\cdots,\nonumber\\
&& x_0=x, \; x_1 = \epsilon x, 
\end{eqnarray}
and solve Eq. (\ref{maxwell_disp_weak_fourier}) order by
order in $\epsilon$. The zero-order in $\epsilon$ yields
\begin{equation} \label{eq_mdf_zero_order}
\left(\frac{\partial^2}{\partial x_0^2}+\omega^2\right)\hat{E_0} = 0
\end{equation}
with the solution
\begin{equation} \label{sol_mdf_zero_order}
\hat{E}_0(x_0,x_1,\omega)=\hat{A_0}(x_1,\omega){\mathrm{e}}^{i\omega x} + \hat{B_0}(x_1,\omega){\mathrm{e}}^{-i\omega x}
\end{equation}
where 
the term containing $\hat{A_0}$ corresponds to a wave moving to the right and $\hat{B_0}$
to a wave moving to the left. 

%
%
Collecting the first order terms in $\epsilon$, we find the equation for $\hat{E_1}$ as
\begin{eqnarray}\label{eq_mdf_first_order}
\left(\frac{\partial^2}{\partial x_0^2}+\omega^2\right)\hat{E_1}&=& -2\frac{\partial}{\partial x_0}\frac{\partial}{\partial x_1}\hat{E}_0 -\omega^2\hat{\chi}^{(1)}(\omega)\hat{E}_0 \nonumber\\
 &-&\frac{\omega^2}{(2\pi)^2} {\mathcal{N}}(E_0).
\end{eqnarray}
%
The question is now how to formulate the solvability condition for
(\ref{eq_mdf_first_order}) taking into account the nonlocal driving
term on the right hand side of the equation. The fundamental solutions
of the linear operator on the l.h.s. of
Eq. (\ref{eq_mdf_first_order}) are again $\{\exp(i\omega x_0),
\exp(-i\omega x_0)\}$. Following the technique of multiple scales, we apply Fredholm alternative for removing possible resonances which arise from the terms in r.h.s. of Eq. (\ref{eq_mdf_first_order}) proportional to
those fundamental solutions. Using the expression (\ref{sol_mdf_zero_order}), the triple product
(omitting the arguments $x_0$ and $x_1$ for simplicity)
$E_0(\omega_1)E_0(\omega_2)E_0(\omega-\omega_1-\omega_2)$ will yield
eight terms.
%
%
In each of those terms, we find all the possible combinations of
$\omega_1$ and $\omega_2$ that lead to the solution $\hat{E}_1$ proportional to either
$\exp(i\omega x_0)$ or $\exp(-i\omega x_0)$. Those combinations are
responsible for resonance and hence, must be removed. From Eq. (\ref{sol_mdf_zero_order}), we find (the primes denote derivatives with respect to the slow variable $x_1$)
\begin{equation}
-2\frac{\partial}{\partial x_0}\frac{\partial}{\partial x_1}\hat{E}_0(x_0,x_1,\omega)=-2i\left(\hat{A}'_0{\mathrm{e}}^{i\omega x_0}+\hat{B}'_0{\mathrm{e}}^{-i\omega x_0}\right).
\end{equation}
Thus, the contributions proportional to $\exp(i\omega x_0)$, $\exp(-i\omega x_0)$ will appear in
the equation for $\hat{A}'_0$ and $\hat{B}'_0$, respectively. Let
us now observe the first of the eight terms (omitting the
evolution variable $x_0$ for the sake of easy typesetting)
\begin{displaymath}
\hat{A}_0(\omega_1)\hat{A}_0(\omega_2)\hat{A}_0(\omega-\omega_1-\omega_2){\mathrm{e}}^{i\omega x_0}.
\end{displaymath}
The oscillations arise from ${\mathrm{e}}^{i\omega x_0}$ and thus, this term will appear {\em only} in the equation for $\hat{A}'_0$. The next term 
\begin{displaymath}
 \hat{A}_0(\omega_1)\hat{A}_0(\omega_2)\hat{B}_0(\omega-\omega_1-\omega_2){\mathrm{e}}^{i(-\omega+2\omega_1+2\omega_2)x_0}
\end{displaymath}
incorporates oscillations with different frequencies corresponding to
%
${\mathrm{e}}^{i(-\omega+2\omega_1+2\omega_2)x_0}$.
%
Contributions to resonance terms occur if
\begin{enumerate}
\item $-\omega+2\omega_1+2\omega_2 = \omega$ responsible for contributions to the equation for $\hat{A}'_0$.
\item $-\omega+2\omega_1+2\omega_2 = -\omega$ responsible for contributions to the equation for $\hat{B}'_0$.
\end{enumerate}
Therefore, the solvability conditions lead to the system of equations for $\hat{A}'_0$ and $\hat{B}'_0$,
%
%
\begin{align}
&-2i\omega \hat{A}'_0(\omega) = \omega^2\hat{\chi}^{(1)}\hat{A}_0(\omega) + \frac{\omega^2}{(2\pi)^2} {\mathcal{N}}(A_0) +\nonumber \\
&\frac{\omega^2}{(2\pi)^2} \left( 3\hat{B}_0(0)\int \hat{\chi}^{(3)}(\omega_1,\omega-\omega_2,0)\hat{A}_0(\omega_1)\hat{A}_0(\omega-\omega_1)d\omega_1 \nonumber \right. \\
&+ \left. 3\hat{A}_0(\omega)\int \hat{\chi}^{(3)}(\omega_1,-\omega_1,\omega)\hat{B}_0(\omega_1)\hat{B}_0(-\omega_1)\;d\omega_1\right), \label{a0}\\
&2i\omega \hat{B}'_0(\omega) = \omega^2\hat{\chi}^{(1)}(\omega)\hat{B}_0(\omega) +\frac{\omega^2}{(2\pi)^2} {\mathcal{N}}(B_0)+ \nonumber \\
& \frac{\omega^2}{(2\pi)^2} \left( 3\hat{A}_0(0)\int \hat{\chi}^{(3)}(\omega_1,\omega-\omega_2,0)\hat{B}_0(\omega_1)\hat{B}_0(\omega-\omega_1)d\omega_1 \nonumber\right. \\
&+ \left. 3\hat{B}_0(\omega)\int \hat{\chi}^{(3)}(\omega_1,-\omega_1,\omega)\hat{A}_0(\omega_1)\hat{A}_0(-\omega_1)\;d\omega_1\right). \label{b0}
\end{align}
%
The resulting system of equations for $\hat{A}_0$ and
$\hat{B}_0$ describes the slow evolution of Eq. (\ref{nonlinwave}). 
These provide a significant simplification of the original equation (\ref{nonlinwave}), which are the {\em{first}} order equations in $x_1$ and, if we take the inverse Fourier transform, first order in $t$ as well. Note that the resulting equations preserve the nonlocal structure, which was our main purpose. This pair of equations can be seen as an analog to the
NLSE valid for broad pulses (without the assumption of a signal centered around a specific carrier frequency) or the SPE valid for
short pulses (without the specific assumption about the profile of linear and nonlinear susceptibility). The only assumption necessary to
obtain the above system was the smallness of the dispersive and the
nonlinear terms. Our derivation also suggests how to proceed in
the case where the dispersive term is not small. In this case, the solutions
of the linear problem will involve a more complicated dispersion
relation which will give rise to modified resonance conditions for the nonlinearity. 

Note that Eqs. (\ref{a0}), (\ref{b0}) also account for coupling between the forward and backward moving waves $A_0$, $B_0$, respectively. The coupling
is weak in the sense that it only consists of integrated quantities.
In some cases, however, e.g. when the susceptibilities have an additional
dependence on the evolution variable $x$ such as in certain photonic crystal
structures, the coupling between $A_0$ and $B_0$ can become important.
%

\section{The dispersion-free case}\label{dispersion-free}

In the search of soliton solutions of the underlying nonlocal equations, we first consider a particular case, absence of linear dispersion $\chi^{(1)}=0$ and constant nonlinear susceptibility $\chi^{(3)}=1$. Then, we take the inverse Fourier transform of (\ref{a0}), (\ref{b0}) to obtain differential equations in time domain. 

We first notice that $\hat{B}'_0(0,x_1)=0$ implies $\hat{B}_0(0)=\hat{B}_0(\omega=0,x_1)$ is constant as a function of $x_1$ and the corresponding statement also holds for $\hat{A}_0(\omega=0,x_1)$. For convenience of notation, let us define
\begin{displaymath}
2\pi B_{\mathrm{zero}} = \hat{B}_0(0), \qquad 2\pi A_{\mathrm{zero}} = 
\hat{A}_0(0).
\end{displaymath}
After straightforward calculation we find that
\begin{eqnarray}
&&A_{{\mathrm{int}}} := \frac{1}{(2\pi)^2}\int \hat{A}_0(\omega_1)\hat{A}_0(-\omega_1)\;d\omega_1, \nonumber\\
&&B_{{\mathrm{int}}}:=\frac{1}{(2\pi)^2}\int \hat{B}_0(\omega_1)\hat{B}_0(-\omega_1)\;d\omega_1
\end{eqnarray}
are constants of motion as well. Therefore, after Fourier transform back in time domain we have ($A_0=A_0(x,t)$ and $B_0=B_0(x,t)$)
\begin{eqnarray}
\frac{\pp}{\pp x_1}A_0+\frac{1}{2}\frac{\pp}{\pp t}\left(A_0^3+3B_{\mathrm{zero}}A_0^2+3B_{\mathrm{int}}A_0\right) = 0, \label{eq_df_A}\\ 
\frac{\pp}{\pp x_1}B_0-\frac{1}{2}\frac{\pp}{\pp t}\left(B_0^3+3A_{\mathrm{zero}}B_0^2+3A_{\mathrm{int}}B_0\right) \label{eq_df_B}= 0.
\end{eqnarray}
As mentioned before, the forward and backward propagating waves $A_0$, $B_0$
are coupled. This coupling, however, is weak in a sense that it occurs
only through the constants of motions. By choosing initial condition
$B_0(x_1=0,t)\equiv 0$ we can eliminate the backward propagating wave
$B_0$ entirely at the leading order and the equation for $A_0$ is
reduced to a Burger's type equation
\begin{equation}
\frac{\pp}{\pp x_1}A_0 + \frac{1}{2}\frac{\pp}{\pp t}A_0^3 = 0,
\end{equation}
with the (implicit) solution
\begin{equation}
A_0(t,x_1) = f\left(t-\frac{3}{2}A_0^2x_1\right).
\end{equation}
Here, the function $f$ is determined by the initial condition of the pulse at $t=0$. For typical pulse profiles (e.g., a Gaussian pulse), we see that shocks will be generated eventually. This was expected as we excluded all the linear dispersion terms from our considerations that could have helped to smoothen the solution.

\section{Solitary waves}\label{dispersion}

In order to regularize the shocks and further obtain stable wave
propagation, we bring back a small dispersion term ($\hat{\chi}^{(3)}$ is still set to be constant for simplicity) and find
\begin{eqnarray} \label{maxwell_small_disp}
\left(\frac{\partial^2}{\partial x^2}-\frac{\partial^2}{\partial t^2}\right)E(x,t)&=&\epsilon \frac{\partial^2}{\partial t^2}\int \chi^{(1)}(t-\tau)E(x,\tau)d\tau \nonumber\\
&+& \epsilon \frac{\partial^2}{\partial t^2} E(x,t)^3.
\end{eqnarray}
Now we apply the same analysis used before and obtain an additional nonlocal term in Eqs. (\ref{eq_df_A}), (\ref{eq_df_B}),
\begin{eqnarray}
\frac{\pp}{\pp x_1}A_0&+&\frac{1}{2}\frac{\pp}{\pp t}\left(A_0^3+3B_{\mathrm{zero}}A_0^2+3B_{\mathrm{int}}A_0 \right.\nonumber\\
&+& \left.\int\chi^{(1)}(t-\tau)A_0(x_1,\tau)d\tau\right) = 0, \label{eq_sd_A}\\ 
\frac{\pp}{\pp x_1}B_0&-&\frac{1}{2}\frac{\pp}{\pp t}\left(B_0^3+3A_{\mathrm{zero}}B_0^2+3A_{\mathrm{int}}B_0 \right.\nonumber\\
&+& \left.\int\chi^{(1)}(t-\tau)B_0(x_1,\tau)d\tau \right) = 0. \label{eq_sd_B}
\end{eqnarray}
%
Again we assume the back-scattered part $B_0$ is suppressed. Motivated by the recent developments e.g., in photonic crystal structures a wide range of susceptibilities can be engineered, we search for the profiles of $A_0$ and $\chi^{(1)}$ which can stabilize the shock formation. One simple approach to finding a stationary solution is to set $\pp A_0/ \pp x_1 = 0$.  This implies we need to find $A_0$ and $\chi^{(1)}$ such that
\begin{equation} \label{cond_stat}
A_0^3 + \int\chi^{(1)}(t-\tau)A_0(x_1,\tau)d\tau = 0.
\end{equation}
Denoting $\hat{\chi}^{(1)}={\mathcal{F}}(\chi^{(1)})$, we rewrite Eq. (\ref{cond_stat}) in Fourier domain,
\begin{equation}
{\mathcal{F}}(A_0^3) + \hat{\chi}^{(1)} \hat{A_0} = 0.
\end{equation}
If we assume a Lorentz profile for $A_0$
\begin{equation} \label{lorentz}
A_0(t)=\frac{\alpha}{1+(\beta t)^2},
\end{equation}
we can solve Eq. (\ref{cond_stat}) using Fourier transform. The Fourier
transform of $A_0$ is
\begin{equation}
\hat{A}_0(\omega) = \pi \frac{\alpha}{\beta} {\mathrm{e}}^{-|\omega|/\beta}
\end{equation}
and, using the residue theorem, we obtain the Fourier transform
of $A_0^3$, 
\begin{equation}
{\mathcal{F}}(A_0^3)(\omega) = \frac{\pi}{8}\frac{\alpha^3}{\beta^3}
{\mathrm{e}}^{-|\omega|/\beta} (\omega^2+3\beta|\omega|+3\beta^2).
\end{equation}
We find that Eq. (\ref{lorentz}) is a stationary solution provided 
that the susceptibility takes the following parabolic form:
\begin{equation} \label{sol_sus}
\hat{\chi}^{(1)}(\omega)=-\frac{1}{8}\frac{\alpha^2}{\beta^2}(\omega^2+3\beta|\omega|+3\beta^2).
\end{equation}
We expect that a pulse of this particular shape, for the given
susceptibility will not undergo changes at the leading order, hence
propagates as a stable nonlinear wave. The dispersion compensates for
the effect of nonlocal nonlinear term not only in a small frequency
range but for all possible values of $\omega$.  Due to this balance
between convolution terms in the nonlinearity and dispersion, we can
obtain a stable solitary wave.

Most of the physical examples for Maxwell's equations, however,
require a modulated initial condition. Therefore, we consider a pulse of the form
\begin{equation} \label{modulated_initialcondition}
\tilde A_0(t)= A_0(t)\,\cos(\omega_0 t)
\end{equation}
which will have its maximal frequency coefficients in the neighborhood
of $\omega_0$. We extend our analysis to this case of a modulated initial condition and find a susceptibility
for which the given pulse will maintain its stable propagation. Since the
Fourier transforms of $\tilde A_0$ and $\tilde A_0^3$ can be found
directly as
\begin{eqnarray}
{\mathcal{F}}(\tilde A_0)&=&\frac{1}{2}\left(\hat A_0(\omega-\omega_0)+
                                           \hat A_0(\omega+\omega_0)\right) \\
{\mathcal{F}}(\tilde A_0^3)&=&\frac{1}{8}\left({\mathcal{F}}(A_0^3)(\omega+3\omega_0) +3{\mathcal{F}}(A_0^3)(\omega+\omega_0)  \right. \nonumber\\ &+& \left. \,3{\mathcal{F}}(A_0^3)(\omega-\omega_0+{\mathcal{F}}(A_0^3)(\omega-3\omega_0)\right)
\end{eqnarray}
we find the susceptibility ${\mathcal{F}}(\tilde \chi^{(1)})$
for the modulated solitary wave as
\begin{equation} \label{modulated_susceptibility}
{\mathcal{F}}(\tilde \chi^{(1)}) = - \frac{{\mathcal{F}}(\tilde A_0^3)}{{\mathcal{F}}(\tilde A_0)}
\end{equation}
It is important to note that the presented susceptibilities
(\ref{sol_sus}) and (\ref{modulated_susceptibility}) do not correspond
to susceptibilities of any known material. Although they need to be engineered
artificially, recent progress in manufacturing particular photonic
crystal structures \cite{Mingaleev-Kivshar:2002,bjarklev-broeng-Bjarklev:2003,hansen-etal:2005} makes it reasonable to expect the possibility to
design such and similar susceptibilities in a not too distant future.
On the other hand, the above analysis offers a variety of different
approaches for regularizing shocks to obtain stable propagation of
nonlinear waves. One possibility, for example, would be to introduce a
dependence of the slow variable $x_1$ to $\chi^{(1)}$ (as in certain
photonic crystal structures). This increases the mathematical
difficulty to find solutions to the system of equations
(\ref{eq_sd_A}), (\ref{eq_sd_B}) but leads to more freedom in the
choice of $A_0$ and $\chi^{(1)}$.

\section{Numerical Results}\label{numerics}

\begin{figure}[htb]
\centering
\includegraphics[height=0.45\textwidth, angle=-90]{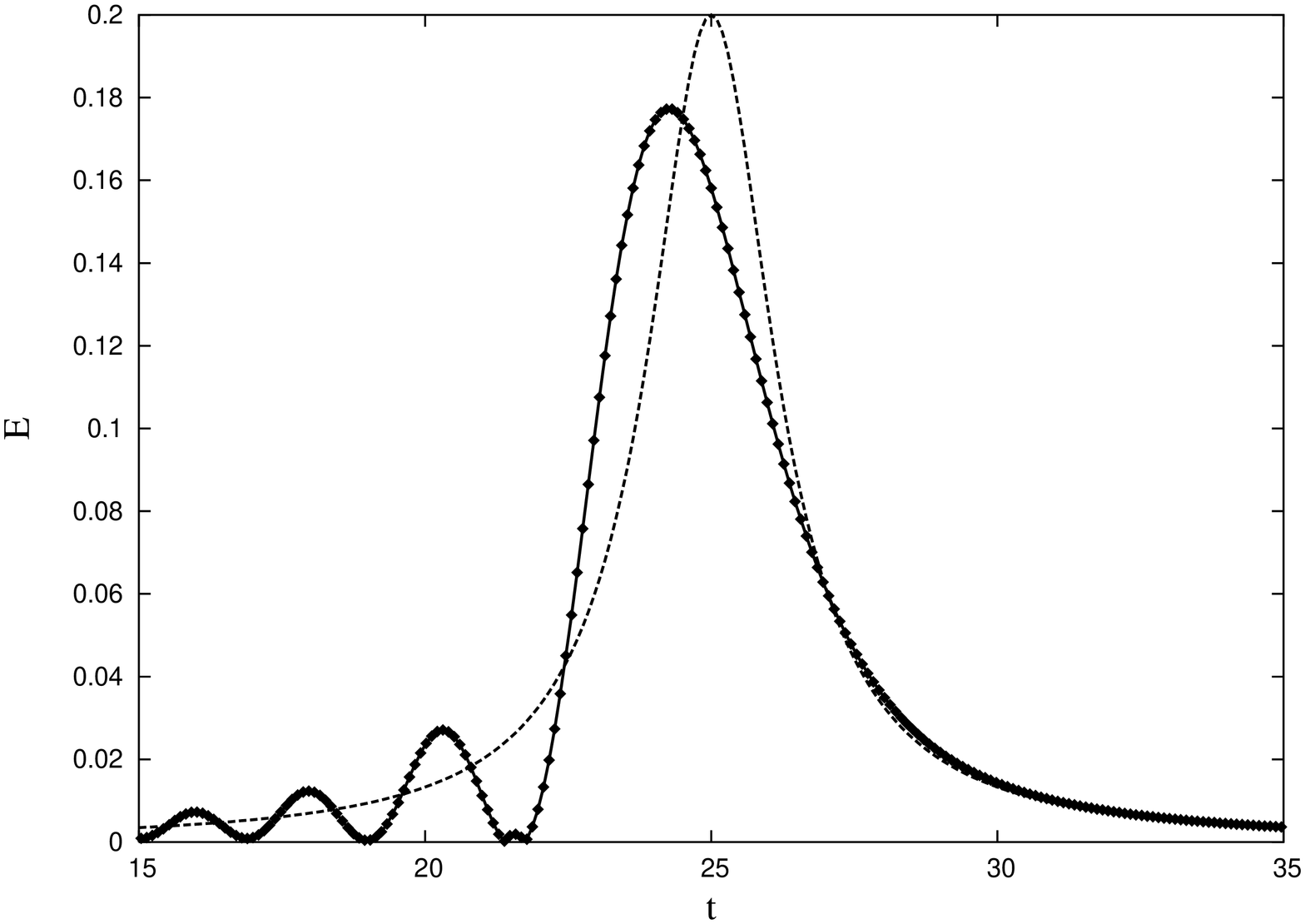}
\hfill
\includegraphics[height=0.45\textwidth, angle=-90]{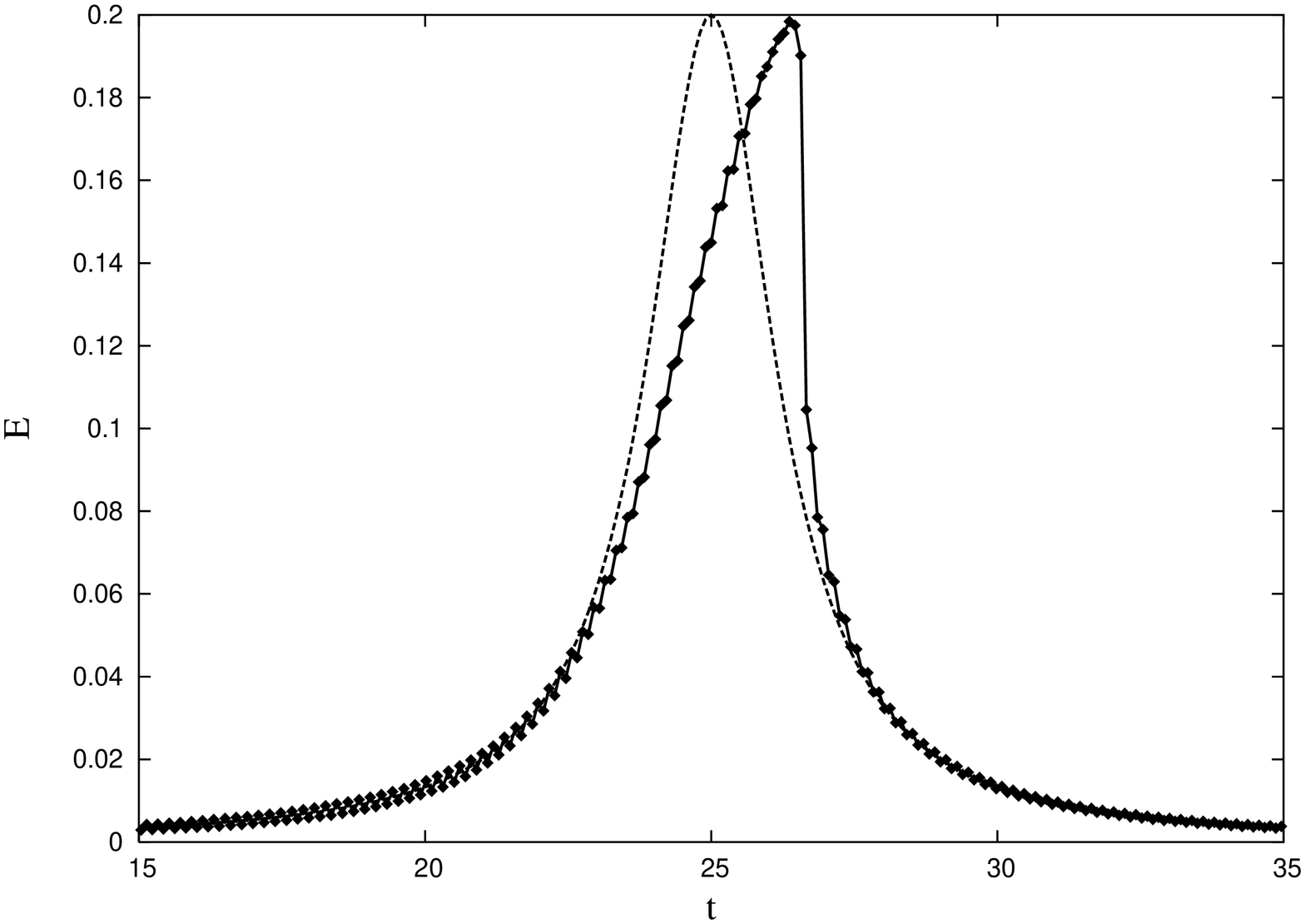}
\caption{{\protect\small Figure on top: The diamonds show the solution of 
Maxwell's wave equation (\ref{nonlinwave}) in the linear case
$\chi^{(3)}=0$ versus the initial pulse (dashes) in a moving
frame. The linear susceptibility $\chi^{(1)}$ leads to dispersion.
Bottom figure: The diamonds represents the solution of Maxwell's wave equation
(\ref{nonlinwave}) in the purely nonlinear case $\chi^{(1)}=0$ versus
the initial pulse (dashes) in a moving frame. It clearly illustrates
the formation of a shock.}}\label{fig:comparison}
\end{figure}

\begin{figure}[htb]
\centering
\includegraphics[height=0.45\textwidth, angle=-90]{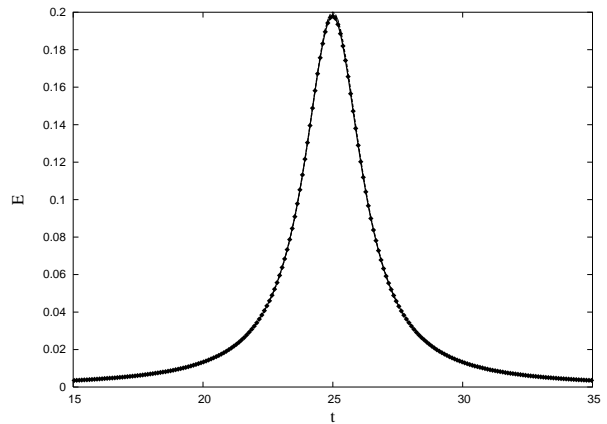}
\caption{{\protect\small Propagation of the solitary wave. The diamonds represent the solution of Maxwell's wave equation with both dispersive and
nonlinear terms. The stability of the pulse propagation is due to the
balance of both effects.}}
\label{fig:soliton}
\end{figure}

In order to show the stability of the solitary wave (\ref{lorentz})
numerically, we use its analytic expression together with the
corresponding susceptibility (\ref{sol_sus}) for simulations of the
nonlocal nonlinear wave equation (\ref{nonlinwave}). Here, we chose
$\alpha=0.2$ and $\beta=0.75$ leading to an effective $\epsilon$ of
around $0.07$.  Transmitting the solitary wave to a distance $x=25$,
which is longer than ${\mathcal{O}}(1/\epsilon)$, enables us to
capture all three important effects: Pulse distortion due to
dispersion, pulse distortion due to nonlinearity, and stable
propagation of the nonlinear soliton due to a nonlocal balance of
dispersive and nonlinear effects . Figure \ref{fig:comparison}
presents the pure linear and nonlinear effects.
In Fig. \ref{fig:soliton}, we present the propagation of the solitary
wave combining the linear and nonlinear effects. Although the
characteristics of the soliton are resulted from a first-order
approximation, the numerical simulations show that the balance of
nonlinear convolution terms and dispersion generates a stable solution
whose shape changes only slightly during the propagation.
%
%

The second numerical simulation shows the stable propagation of the modulated initial pulse given by (\ref{modulated_initialcondition}) and corresponding susceptibility (\ref{modulated_susceptibility}).  Here, we chose simulation parameters $\alpha=0.2$ and $\beta=1.0$ and $\omega_0=5.0$. The allowed frequency range was restricted to frequencies below $2\,\omega_0$ to suppress generation of third harmonics.  
\begin{figure}[htb]
\centering
\includegraphics[height=0.45\textwidth, angle=-90]{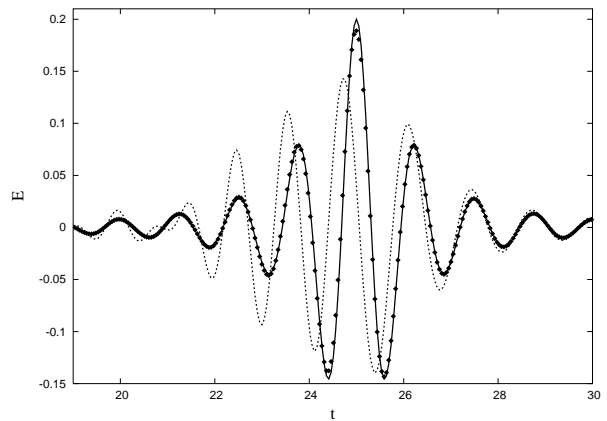}
\caption{{\protect\small Propagation of the modulated solitary wave. The diamonds represent the solution of Maxwell's wave equation with both dispersive and
nonlinear terms, the continuous line the initial condition in a moving
frame. The dashed line is the linear solution of Maxwell's equations
with the same initial condition. The figure shows how nonlinearity allows
the solitary wave to maintain its initial shape.}}
\label{fig:modulated_soliton}
\end{figure}
Figure \ref{fig:modulated_soliton} shows how well nonlinearity and
dispersion are balanced in this case. The pulse experiences
only very small deviations from its original shape that are due to
higher-order effects.  From this figure it is also clear that this
solitary wave represents an ultrashort pulse, hence does not fall into
the regime of the cubic nonlinear Schr{\"o}dinger equation. It is a new type of ultra short solitary wave that can exist due to the balance
of dispersion and nonlinearity in a very broad frequency range.

\section{Conclusion} \label{summary}

We derived a slowly varying envelope equation for ultra-short pulses
from the cubic nonlinear Maxwell's equation. The derivation was
carried out without imposing any locality assumption unlike NLSE,
which allows to preserve the phenomena that are due to the retarded
response of the media in ultra-short pulse dynamics. We also did not
assume any specific types of material susceptibility and hence, our
new model can be applied more generally than the SPE. We showed that
for certain susceptibility of material the resulting approximative
equations possess a solitary wave solution due to a (nonlocal) balance
between dispersion and nonlinearity. The stable propagation of this
solitary wave is confirmed by numerical simulations based on the full
Maxwell's equation.

\section*{Acknowledgments}
We are grateful to A. Aceves for valuable comments and useful discussions. 
The work of T. Sch{\"a}fer was supported by CUNY research foundation
through the grant PSCOOC-36-176. The work of Y. Chung was supported by National Science Foundation through the grant NSF-DMS-0505618.

\bibliography{master}

\begin{thebibliography}{10}

\bibitem{bwzw99}
E.~B. Brown, E.~Wu, W.~Zipfel, and W.~W. Webb.
\newblock Measurement of molecular diffusion in solution by multiphoton
  fluorescence photobleaching recovery.
\newblock {\em Biophys. Jour.}, 77:2837--2849, 1999.

\bibitem{ktrs00}
M.~W. Kimmel, R.~Trebino, J.~Ranka, and A.~J. Stentz.
\newblock {\em CLEO 2000, CFL7, San Francisco}.

\bibitem{k98}
J.~C. Knight, J.~Broeng, T.~A. Birks, and P.~St.~J. Russell.
\newblock Photonic band gap guidance in optical fibers.
\newblock {\em Science}, 282:1476--1478, 1998.

\bibitem{rothenberg:1992}
J.~E. Rothenberg.
\newblock Space-time focusing: breakdown of the slowly varying envelope
  approximation in the self-focusing of femtosecond pulses.
\newblock {\em Opt. Lett.}, 17:1340--1342, 1992.

\bibitem{husakou}
A.~V. Husakou and J.~Herrmann.
\newblock Supercontinuum generation of higher-order solitons by fission in
  photonic crystal fibers.
\newblock {\em Phys. Rev. Lett.}, 87:203901, 2001.

\bibitem{hgz02}
J.~Herrmann, U.~Griebner, N.~Zhavoronkov, A.~Husakou, D.~Nickel, J.~C. Knight,
  and W.~J. Wadsworth.
\newblock Experimental evidence for supercontinuum generation by fission of
  higher-order solitons in photonic crystal fibers.
\newblock {\em Phys. Rev. Lett.}, 88:173901, 2002.

\bibitem{dudley}
J.~M. Dudley, X.~Gu, L.~Xu, M.~Kimmel, E.~Zeek, P.~O'Shea, R.~Trebinoand~S.
  Coen, and R.~S. Windeler.
\newblock Cross-correlation frequency resolved optical gating analysis of
  broadband continuum generation in photonic crystal fiber: simulations and
  experiments.
\newblock {\em Opt. Exp.}, 10:1215--1221, 2002.

\bibitem{fio}
F.~G. Omenetto, A.~J. Taylor, M.~D. Moores, J.~Arriaga, J.~C. Knight, W.~J.
  Wadsworth, and P.~St. Russell.
\newblock Simultaneous generation of spectrally distinct third harmonics in a
  photonic crystal fiber.
\newblock {\em Opt. Lett.}, 26:1158--1160, 2001.

\bibitem{schaefer-wayne:2004}
T.~Sch{\"a}fer and C.~E. Wayne.
\newblock Propagation of ultra-short optical pulses in cubic nonlinear media.
\newblock {\em Physica D}, 196:90--105, 2004.

\bibitem{chung-jones-etal:2005}
Y.~Chung, C.K.R.T. Jones, T.~Sch{\"a}fer, and C.~E. Wayne.
\newblock Ultra-short pulses in linear and nonlinear media.
\newblock {\em Nonlinearity}, 18:1351--1374, 2005.

\bibitem{boyd:1992}
R.~W. Boyd.
\newblock {\em Nonlinear Optics}.
\newblock Academic Press, Boston, 1992.

\bibitem{nikolov-neshev-etal:2003}
N.~I. Nikolov, D.~Neshev, O.~Bang, and W.~Z. Krolikowski.
\newblock Quadratic solitons as nonlocal solitons.
\newblock {\em Phys Rev E}, 68:036614, 2003.

\bibitem{holmes:1995}
M.~H. Holmes.
\newblock {\em Introduction to Perturbation Methods}.
\newblock Springer, New York, 1995.

\bibitem{Mingaleev-Kivshar:2002}
S.~Mingaleev and Y.~Kivshar.
\newblock Nonlinear photonic crystals; toward all-optical technologies.
\newblock {\em Optics $\&$ Photonics News}, 13, July 2002.

\bibitem{bjarklev-broeng-Bjarklev:2003}
A.~Bjarklev, J.~Broeng, and A.~S. Bjarklev.
\newblock {\em Photonic Cryatal Fibers}.
\newblock Kluwer Academic Publishers, Norwell, Boston, 2003.

\bibitem{hansen-etal:2005}
K.~P. Hansen, J.~Broeng, P.~M.~W. Skovgaard, J.~R. Folkenberg, M.~D. Nielsen,
  A.~Peterson, T.~P. Hansen, C.~Jakobsen, H.~R. Simonsen, J.~Limpert, and
  F.~Salin.
\newblock High-power photonic crystal fiber lasers: Design, handling and
  subassemblies.
\newblock {\em Photonics West}, San Jose, CA, 2005.

\end{thebibliography}

\end{document}